\documentclass{Interspeech2024}
\usepackage{multirow,array}
\usepackage{cite}
\usepackage{subfig}
\usepackage{graphicx}
\usepackage{booktabs}

\interspeechcameraready

\title{How Should We Extract Discrete Audio Tokens from Self-Supervised Models?\thanks{Accepted at Interspeech 2024}}

\name[affiliation={1,2}]{Pooneh}{Mousavi}
\name[affiliation={3}]{Jarod}{Duret}
\name[affiliation={4}]{Salah}{Zaiem}
\name[affiliation={1,2}]{Luca}{Della Libera}
\name[affiliation={5,2}]{Artem}{Ploujnikov}
\name[affiliation={6,2,1}]{Cem}{Subakan}
\name[affiliation={1,2,5}]{Mirco}{Ravanelli}
\address{
  $^1$Concordia University, Canada
  $^2$Mila - Quebec AI Institute, Canada 
  $^3$Avignon Université, France 
  $^4$Telecom Paris, France 
  $^5$Université de Montréal, Canada 
  $^6$Université Laval, Canada}
\email{pooneh.mousavi@mail.concordia.ca}


\begin{document}

\maketitle

\begin{abstract}
% Mirco's version (only minor modifications)
Discrete audio tokens have recently gained attention for their potential to bridge the gap between audio and language processing. Ideal audio tokens must preserve content, paralinguistic elements, speaker identity, and many other audio details. Current audio tokenization methods fall into two categories: Semantic tokens, acquired through quantization of Self-Supervised Learning (SSL) models, and Neural compression-based tokens (codecs). Although previous studies have benchmarked codec models to identify optimal configurations, the ideal setup for quantizing pretrained SSL models remains unclear.

This paper explores the optimal configuration of semantic tokens across discriminative and generative tasks. We propose a scalable solution to train a universal vocoder across multiple SSL layers. Furthermore, an attention mechanism is employed to identify task-specific influential layers, enhancing the adaptability and performance of semantic tokens in diverse audio applications.
%The effectiveness of acoustic models depends on the provided features, with a recent focus on discretizing audio signals. Desired sound tokens must effectively preserve content, paralinguistic elements, speaker identity, and intricate audio details. Current audio tokenization methods fall into two categories: Semantic tokens, acquired through quantization of Self-Supervised Learning (SSL) models, and Neural compression-based tokens. Despite studies benchmarking codec models for optimal configurations, the ideal setup for preserving sound information through semantic tokens remains unanswered. This paper explores the optimal configuration of semantic tokens across discriminative and generative tasks. We propose a bitrate scalability solution to train a universal vocoder across multiple SSL layers. Furthermore, an attention mechanism is employed to identify task-specific influential layers, enhancing the adaptability and performance of semantic tokens in diverse audio applications.
\end{abstract}

\section{Introduction}

% Mirco's version (a bit too long)
Learning effective, efficient, and robust representations is a core problem in modern audio and speech processing systems \cite{lecun2015deep}.
Over the past few years, continuous representations learned by large self-supervised models such as Wav2Vec2 \cite{baevski2020wav2vec}, WavLM \cite{chen2022wavlm}, and HuBERT \cite{hsu2021hubert} have achieved unprecedented performance.
A recent research trend consists of learning discrete audio representations instead of continuous ones, resulting in what is known as \textit{audio tokens}. These discrete tokens offer several potential advantages. Firstly, they facilitate the development of audio language models (LMs)~\cite{borsos2023audiolm,rubenstein2023audiopalm,chen2023lauragpt,wang2023viola,wang2301neural,wang2023speechx} and the creation of multi-modal large language models~\cite{geminiteam2023gemini}, which can emit audio, text, and visual tokens. Additionally, their compression potential can contribute to efficient data transmission and storage. Discrete tokens also enable us to address audio generation tasks such as speech enhancement and synthesis using classification methods, instead of relying on complex high-dimensional regression models. %Handling these tasks as classification rather than complex high-dimensional regression problems is notably easier with current deep learning techniques~\cite{GoodBengCour16}.

%Recently, discrete audio tokens have attracted attention for thei capacity to transform continuous audio into discrete tokens. These tokens serve a dual purpose, aiding in the development of audio language models (LMs)~\cite{borsos2023audiolm,rubenstein2023audiopalm,chen2023lauragpt,wang2023viola,wang2301neural,wang2023speechx,agostinelli2023musiclm,kreuk2022audiogen,copet2024simple,yang2023uniaudio} by bridging the gap between audio and language processing. Simultaneously, they contribute to the reduction of data transmission latency to enhance computational efficiency without compromising performance.
Following the terminology from~\cite{borsos2023audiolm,zhang2023speechtokenizer}, audio tokenization techniques can be broadly categorized into Compression-based (codecs) tokens and Semantic tokens.
Compression-based tokens~\cite{zeghidour2021soundstream, defossez2022high, kumar2024high, yang2023hifi} utilize encoder-decoder architectures coupled with Residual Vector Quantization (RVQ)~\cite{zeghidour2021soundstream}. They are explicitly trained to accurately reconstruct the original audio, making them particularly suitable for audio generation tasks.
Semantic tokens~\cite{polyak2021speech, wells2022phonetic, chung2021w2v}, on the other hand, are generated through clustering or quantization of the layers of Self-Supervised Learning (SSL) models~\cite{baevski2020wav2vec, chen2022wavlm, hsu2021hubert}. Often, this involves selecting a layer from the pretrained SSL model and clustering its representations, typically with the k-means algorithm. Semantic tokens primarily capture coarse information such as phonetic, semantics, and syntactic details. Since they are not explicitly trained to achieve accurate waveform reconstruction, it is more natural to use them in discriminative tasks like Automatic Speech Recognition (ASR). Recent research, however, has shown that semantic tokens can be effective for generative tasks as well~\cite{wang2023selm, yang2023towards}. Additionally, semantic tokens have been used in a hybrid tokenizer~\cite{zhang2023speechtokenizer,du2023funcodec}. This hybrid approach combines semantic and compression-based tokens, separating content information in the initial layer while preserving paralinguistic details in subsequent layers. A similar strategy has been widely adopted in audio LLMs ~\cite{borsos2023audiolm, rubenstein2023audiopalm, chen2023lauragpt}.
%Semantic tokens are particularly intriguing as they provide "pseudo-text" useful for training text-free speech language models ~\cite{borsos2023audiolm, rubenstein2023audiopalm, chen2023lauragpt, agostinelli2023musiclm, lakhotia2021generative, kharitonov2021text, nguyen2023generative, popuri2022enhanced, inaguma2022unity, wang2023selm}  opening up the possibility of leveraging established Natural Language Processing (NLP) techniques, such as prompt tuning~\cite{chang2022speechprompt, chang2023speechprompt, wu2023speechgen}, in-context learning~\cite{hsu2023exploration}, and instruction tuning~\cite{kuan2023towards, huang2023dynamic}, for diverse speech-processing tasks. 
%While prior research has mainly focussed on benchmarking optimal configurations for codec models~\cite{wu2024codec, puvvada2023discrete}, 
Nevertheless, the most effective setting for extracting semantic tokens remains largely unclear. Recent studies have primarily focused on ASR and Speech Translation~\cite{chang2023exploring, zhang2023dub, chang2023exploration}, without considering a broader range of discriminative and generative tasks.

This paper addresses this gap by evaluating the effects of different heuristics required to derive semantic tokens for several discriminative and generative tasks, such as speech recognition, speaker recognition, emotion classification, speech enhancement, and text-to-speech. We investigate various crucial aspects, including the impact of the number of clusters and the selection of the intermediate layer of the SSL model to discretize. The latter factor turned out to be crucial and task-dependent, as early layers capture low-level information and higher layers encode content and semantic nuances. Common strategies include using the middle layer~\cite{wang2023selm, polyak2021speech} or leveraging the last layer~\cite{chang2023exploration}. Instead of relying on partial information only, we introduced a novel technique based on an informed layer selection mechanism. We propose to cluster all layers and inject their information into the acoustic models using learnable attention weights. This approach significantly boosts performance while also providing valuable insights into the importance of each layer.

% The selection of layers to discretize is crucial for generative tasks like enhancement and text-to-speech as well.
Since there is no built-in decoder in semantic tokens, a vocoder model for converting the semantic tokens into audio must be trained ~\cite{kong2020hifigan, prenger2018waveglow}. Training such a vocoder is computationally demanding, making it highly impractical to train a separate vocoder for each layer or combination of layers. To address this challenge, we propose a novel scalable vocoder capable of operating with various layer combinations at no additional cost. This is achieved through a layer dropout training scheme, inspired by the bitrate scalability mechanism used in SoundStream ~\cite{zeghidour2021soundstream}. Interestingly, our results show that the scalable vocoder outperforms all vocoders trained on every specific layer.
Finally, for a comprehensive comparison, we provide experimental evidence using both in-domain and out-of-domain datasets for training k-means. For reproducibility and to encourage further research, we release the code, built on the popular SpeechBrain~\cite{speechbrain_ravanelli} toolkit, and pretrained models publicly\footnote{\url{github.com/speechbrain/benchmarks/tree/DASB}}.

\section{Model Design}
\begin{figure}[t]
  \centering
  \includegraphics[width=0.45\textwidth]{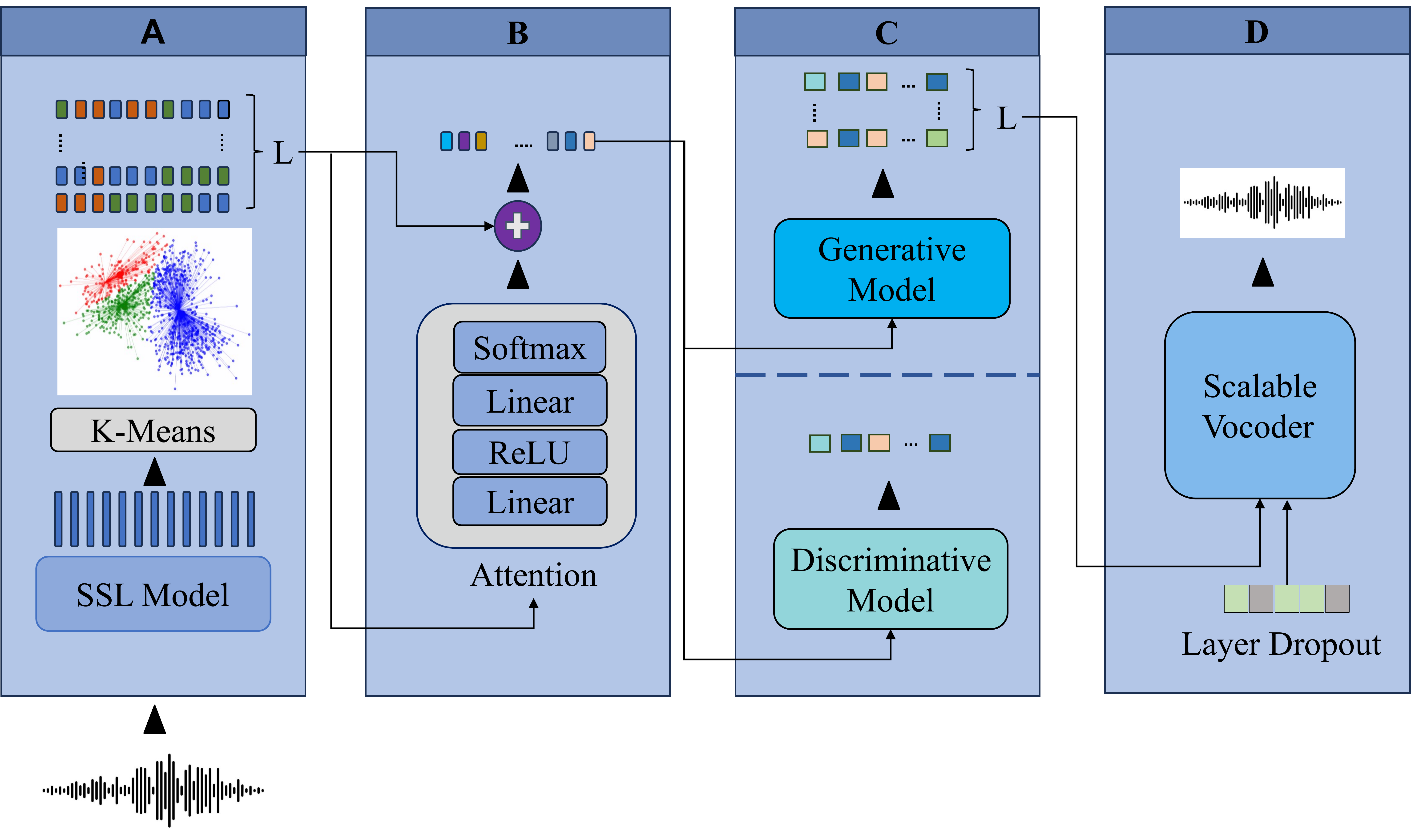}
 \caption{The proposed method for audio token extraction from SSL models: (A) k-means discretizes the continuous representations of each layer, (B) an attention mechanism merges the discrete layer representations, (C) the mixed representations train acoustic models for discriminative and generative tasks, (D) our scalable vocoder generates waveforms (if needed).}
  \label{fig:arch}
\end{figure}

The proposed architecture, illustrated in Fig. \ref{fig:arch}, consists of four components
\begin{enumerate*}[label={\alph*)}]
\item Tokenizer,
\item Informed Layer Selector,
\item Acoustic Model, and
\item Scalable Vocoder
\end{enumerate*}.
The following subsections will describe each module.
% The tokenizer extracts discrete representations from raw audio, while the second step assigns weights to each SSL layer. The final two steps are dedicated to training a downstream task, which may or may not require using the vocoder. 

% \footnote{\url{huggingface.co/microsoft/wavlm-large}}
% \footnote{\url{huggingface.co/facebook/hubert-large-ll60k}}
\subsection{Tokenizer}
For quantization, we cluster five layers taken from two pretrained SSL models using the k-means algorithm independently for each layer. We consider two widely-used models: 
WavLM-large\footnote{\url{huggingface.co/microsoft/wavlm-large}} and HuBERT-large\footnote{\url{huggingface.co/facebook/hubert-large-ll60k}}, both having 24 layers. We choose two layers from the lower part (3, 7) to capture fine-grained information, the middle layer (12), and two layers from the higher part (18, 23) for encoding content and meaning. This selection is based on observation from prior research \cite{chen2022wavlm,yang2021superb} which studied the contribution patterns of different layers across various tasks.
As a result, this set of discrete hierarchical tokens captures rich information from the original audio signal.
Each of the K clusters is assigned a unique index. Additionally, we store the continuous coordinates of each centroid for studying the effect of initializing input embeddings in downstream acoustic models (Sec. \ref{subsec:init}). The outcome of this tokenization process is a tensor $\mathbf{d}$ of shape $B \times T \times n_l$, where B represents the batch size, T is the sequence length, and $n_l$ is the number of discretized layers.

%The pre-trained hierarchical multi-layer tokenizer generates multiple tokens for each time step (one for each selected SSL model layer), resulting in a tensor shape of $ B \times T \times n_l$, where B denotes the batch size, T represents the sequence length, and $n_l$ is the size of the subset chosen from $N_l=[3, 7, 12, 18, 23]$.
% For quantization, we employ a K-Means model trained on five layers extracted from two SSL models, WavLM-large \footnote{https://https://huggingface.co/microsoft/wavlm-large} and Hubert-large \footnote{https://huggingface.co/facebook/hubert-large-ll60k}, each with 24 layers. The selected layers include two from the lower part (3,7) for capturing fine-grained information, two from the higher part (18,23) for encoding content and semantic details, and a middle layer (12). The process involves converting continuous outputs from $n_l$ layers chosen from $N_l$=$[3, 5, 12, 18, 23]$ into discrete tokens using pre-trained K-Means, resulting in an output shape of B $\times$ T $\times$ $n_l$, where B is the batch size, T is the sequence length, and $n_l$ is the number of layers used.

\subsection{Informed Layer Selector}
\label{subsec:informed_layer_selector}
As evident from the SSL literature \cite{zaiem2023speech, yang2021superb, zaiem2023icassp}, the choice of the layer within the SSL model significantly influences the performance of the downstream task of interest.
This decision is equally critical for semantic tokens. Unlike prior methods that rely on heuristic layer selection ~\cite{wang2023selm, polyak2021speech, chang2023exploration}, we integrate the information from our hierarchical multi-layer audio tokens with an attention mechanism.
The attention mechanisms comprise a straightforward multi-layer perceptron (MLP) fed by the embeddings of the audio tokens from each layer. The MLP generates a score for each selected layer, that is normalized by a softmax function as shown in the following equations:

\begin{equation}
z_{l,t} = f\big(emb(d_{l,t})\big) \\
\end{equation}
\begin{equation}
a_{l,t} = \frac{exp(z_{l,t})}{\sum_{k=1}^{nl} exp(z_{k,t})}, \quad h_t = \sum_l a_{l, t} z_{l, t},
\end{equation}
where, $z_{l,t}$ represents the score assigned to layer $l$ at time $t$ by the MLP function $f$. The variable $emb$ refers to the lookup table that assigns embeddings to discrete tokens in $d_l$. The variable $a_{l,t}$ denotes the attention assigned to layer $l$ at time $t$, and lastly $h_t$ is the representation that is fed to the downstream MLP model.
Note that we learn different layer combinations at each time-step, making this mechanism particularly effective.

This simple yet effective approach offers several advantages. Firstly, it enhances flexibility by reducing reliance on heuristic layer selection. The model can now dynamically capture information from different layers for each task.
Additionally, as shown in Sec.\ref{sec:results}, this mechanism yields performance improvements when compared with models utilizing information from a single SSL layer. Lastly, the informed layer selections enhance interpretability, enabling us to analyze the learned weights and understand the relative importance of each layer for each downstream task.
%Unlike previous methods that heuristically choose one layer,  we employ a hierarchical multi-layer tokenizer and an attention layer to prioritize the significance of layers for each task. This approach reduces the dependence on heuristic layer selection, introducing interpretability and incorporating weighted information from various layers to improve overall performance.
% Choosing the right layer from the SSL model is vital for semantic tokens. Lower layers capture detailed information, while higher layers encode content and meaning. Unlike previous methods that choose the middle or last layer, or use canonical correlation analysis (CCA), we employ an attention layer to prioritize the importance of layers for each task. This method reduces reliance on heuristic layer selection, adding interpretability and integrating weighted information from different layers to enhance performance.
\subsection{Acoustic Model}
The mixed representations are fed to a neural model trained to address various downstream tasks\footnote{We train the attention mechanism, embeddings, and the acoustic models jointly.}. While previous studies~\cite{chang2023exploring, zhang2023dub, chang2023exploration} have primarily focused on a few discriminative tasks, we aim to provide evidence across a diverse range of speech applications, considering both discriminative and generative tasks.
We consider ASR, speaker identification, and emotion recognition as discriminative tasks. For generative tasks, we focus on text-to-speech and speech enhancement. The details for each task are reported in Sec. \ref{subsec:experiment}.
% We choose a diverse range of speech tasks, covering both discriminative and generative applications. In discriminative tasks, we explore Automatic Speech Recognition, along with Speaker Identification and Emotion Recognition. For generative tasks, we focus on Text-To-Speech and Speech Enhancement tasks. To adapt the architecture of the model with the multi-layer tokenizer, distinct embeddings are learned for each codebook and combined using the attention obtained in the previous step. The detailed specifications implemented for each task are discussed in section \ref{subsec:experiment}.
% The acoustic model calculates the weighted sum of embeddings from discretized tokens across various layers, using attention from earlier steps, to produce the desired output. We choose a diverse range of speech tasks, covering both discriminative and generative applications. In discriminative tasks, we explore Automatic Speech Recognition in both English and multilingual contexts, along with Speaker Identification and Emotion Recognition. For generative tasks, we focus on Text-to-Speech and enhancement tasks. Section \ref{sec:exp} provides specific details about the models and settings applied to each task.
\subsection{Scalable Vocoder}
Although SSL models such as Wav2vec2, HuBERT, and WavLM are not designed for accurate waveform reconstruction, we can potentially adapt them for generative tasks by training a vocoder on top of their representations.
The dominant approach involves training a separate vocoder for each possible layer combination. However, this approach is impractical and computationally demanding since each downstream task may require a different set of layers.
In this work, we propose a universal and scalable vocoder capable of accommodating various layer combinations. To train such a model, we modify HiFi-GAN \cite{yang2023hifi} to accept a variable number of multi-layer discrete tokens as input. We introduce a layer dropout mechanism, similar to structured dropout \cite{srivastava2014dropout}.
For each input example, we randomly sample $k$ layers from the range $[1, n_l]$,  as shown in the following equations:

\begin{equation}
\mathbf{d}_S \sim \text{Sample}(\mathbf{d}, k) , \quad \mathbf{o} = V(\mathbf{d}_S),\\
\end{equation}
% \begin{equation}

% \end{equation}
where `Sample($\cdot$)' randomly selects \( k \) layers from the discrete representations \( \mathbf{d} \), and \( V \) represents the vocoder function that outputs the waveform \( \mathbf{o} \).
Layers are combined with an attention mechanism that assigns weights to different layers and ensures that the dimensionality of the embeddings remains consistent regardless of the number of layers. 
The model is trained to decode audio by considering all possible combinations of layers.
During inference, the desired combination of layers can be selected.
In addition to its flexibility, this vocoder has demonstrated superior performance compared to vocoders trained on single layers, as we will show in Sec. \ref{sec:results}.

%For generative tasks, the acoustic model's output is fed into the vocoder, converting discrete tokens into audio. Typically, a distinct vocoder model is required for each potential layer combination. However, it is more practical to have a single scalable vocoder capable of working with various layer combinations, reducing the memory needed for storing model parameters. To train such a model, we modify the HiFi-GAN neural vocoder \cite{yang2023hifi} to accept multi-layer discrete tokens as input. We introduce a layer dropout mechanism, akin to structured dropout \cite{srivastava2014dropout}. For each input example, we randomly sample $n_l$ layers from the range $[1, N_l]$, and using only the corresponding $n_l$ layers. Consequently, the model is trained to decode audio for all layer combinations. During inference, the value of $n_l$ can be selected. We utilize the same attention mechanism to obtain the weighted sum of layers, ensuring that the dimensionality of the embeddings remains consistent regardless of the chosen layer number. Therefore, no architectural changes are necessary to accommodate different layers.

\section{Experiments}
\label{subsec:experiment}
% In the following, we describe the discriminative and generative tasks considered in our experiments.
The tasks in our experiments are divided into two groups: Discriminative tasks involving transcription and classification, and generative tasks producing audio.
For the downstream architecture choices and training procedures, we follow the best-performing approaches for classic continuous self-supervised representations~\cite{zaiem2023speech}. We employ 1000 centroids across all tasks, except for ASR and emotion recognition, where we adopt 2000 centroids based on insights from prior research on ASR with discrete representations~\cite{chang2023exploration}. The effect of this selection is probed in Sec.~\ref{subsec:cluster}.
%We divided the set of considered tasks into two sub-groups: discriminative and generative tasks. This division is based on the output of the task: transcription and classification for the former group and audio for the latter. 

\begin{figure}[!t]
    \centering
    \begin{subfloat}{}
    \includegraphics[width=3in]{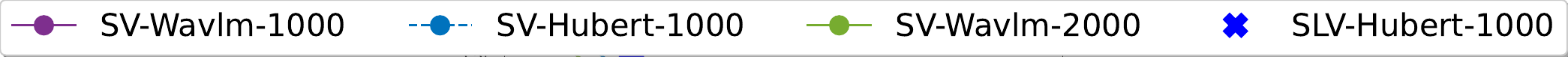}
    \end{subfloat}
    \\
    \begin{subfloat}{}
    \includegraphics[width=1.5in]{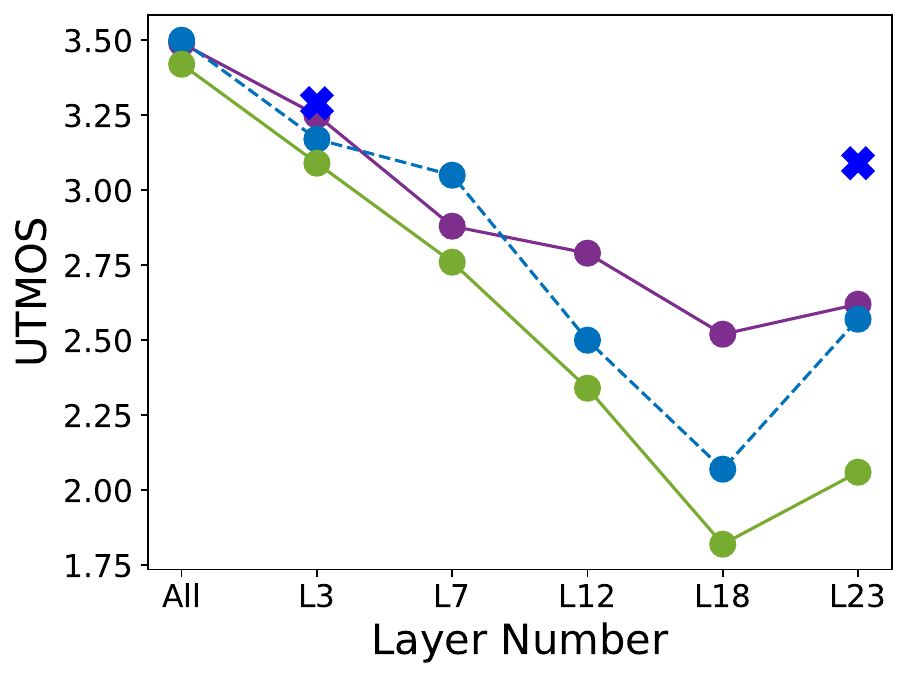}
    \end{subfloat}
    \begin{subfloat}{}
    \includegraphics[width=1.5in]{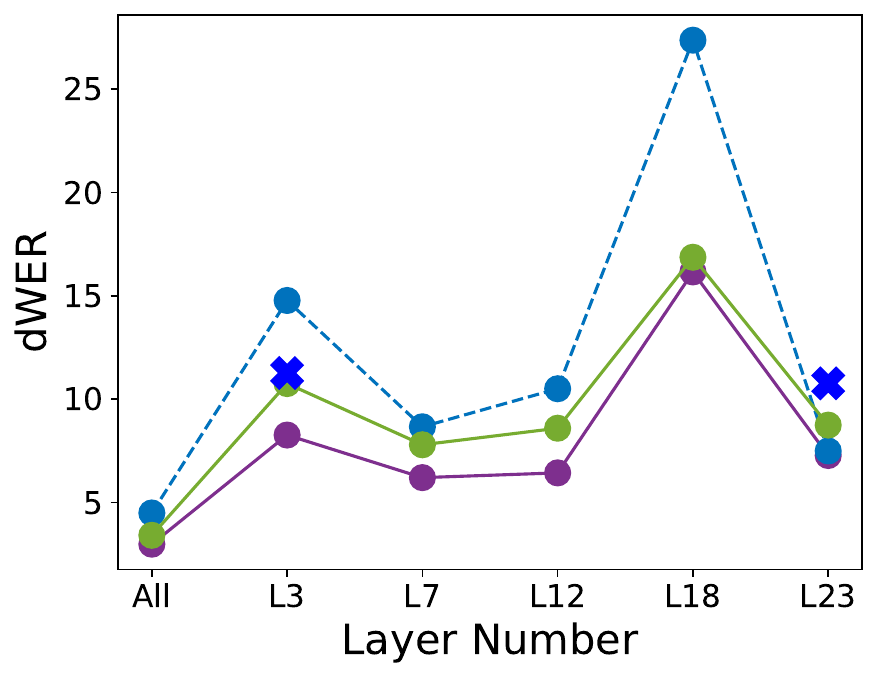}
    \end{subfloat}
\caption{Performance of the Scalable Vocoder (SV) at different layers compared to a Single-Layer Vocoder (SLV). Vocoders and tokenizers are trained using the LJSpeech dataset with 1000 and 2000 centroids.}
\label{fig:imbalance}
\end{figure}

\subsection{Discriminative Tasks}
\noindent  \textbf{Automatic Speech Recognition (ASR)}:   We consider two CTC-based speech recognition tasks. The first one is English ASR using Librispeech \textit{train-clean-100} for training and \textit{test-clean}, \textit{test-other} for testing. The second one uses French data coming from the CommonVoice (CV) 16.1 Corpus \cite{ardila-etal-2020-common}. We select $100$ hours for training,  keeping the original validation and test sets. We use two layers of BiLSTM as a downstream head. The evaluation metric is the Word Error Rate (WER).

\noindent \textbf{Speaker Identification (SID)}: We employ an ECAPA-TDNN model \cite{desplanques2020ecapa}  to determine the speaker identity of each utterance. The widely used VoxCeleb1 \cite{nagrani2017voxceleb} is adopted, and the evaluation metric is accuracy (ACC).

\noindent \textbf{Emotion Recognition (ER)}:  We use  ECAPA-TDNN for emotion recognition \cite{busso2008iemocap} on the IEMOCAP dataset. The task consists of predicting one of the four considered classes: \textit{happy}, \textit{sad}, \textit{angry}, and \textit{neutral}. The evaluation metric is accuracy (ACC).

\subsection{Generative  Tasks}

\noindent \textbf{Speech Enhancement (SE)}: We utilize a non-autoregressive transformer encoder~\cite{transformer}, which consists of 6 layers, 4 attention heads, a model dimension of 256, and a feed-forward layer dimension of 2048. Input tokens are extracted from the noisy signal, and target tokens from the clean one. Training is conducted end-to-end using cross-entropy loss. 
Noisy samples are generated by mixing clean samples from  LJSpeech~\cite{ljspeech17} with noise from WHAM!~\cite{wichern2019wham}. The signal-to-noise ratios (SNRs) are uniformly distributed between 0 and 5 dB. Due to the misalignment of the vocoder's output with the target at the sample level, metrics like Si-SNR can be degraded. Therefore, we use the deep noise suppression mean opinion score (DNSMOS)~\cite{reddy2022dnsmos} for the speech quality metric, following a previous study~\cite{wang2023selm}.
% Speech quality is assessed using the deep noise suppression mean opinion score (DNSMOS)~\cite{reddy2022dnsmos}.
Intelligibility is evaluated through the differential word error rate (dWER)~\cite{wang2021dwer}, which measures the WER between the transcribed enhanced signal and the transcribed target signal. Transcriptions are obtained using the small version of Whisper~\cite{radford2022robust}.

\noindent \textbf{Text-to-Speech (TTS)}: We train an end-to-end autoregressive Transformer\cite{transformer} with 6 layers in the encoder, 12 layers in the decoder, 4 attention heads, a model dimension of 512, and a feed-forward layer in 2048. To facilitate convergence, we employ guided attention~\cite{guided-attention}. The model takes text embeddings as its input and generates the audio tokens for each considered layer.
We utilize a shared transformer decoder, where each tokenizer head has its own learned embedding, and there is a distinct final linear layer for each token. We train all models on the LJSpeech dataset~\cite{ljspeech17}. For assessing speech quality, we use UTMOS ~\cite{1360861705599880960} to estimate human quality ratings. To evaluate fidelity to the text, we assess generated samples using the WER computed with the small version of Whisper~\cite{radford2022robust}.

% For the Text-to-Speech task, we train a simple end-to-end autoregressive Transformer\cite{transformer} model with text embeddings used as inputs and audio tokens obtained from self-supervised models as outputs. In order to facilitate convergence, we employ a guided attention mechanism similar to the one proposed in Tachibana et al\cite{guided-attention}. Unlike a vanilla sequence-to-sequence transformer, the audio signal encoded using the hierarchical multi-layer tokenizer requires the model to output multiple tokens at every time step. We achieve this using a shared transformer decoder with a separate embedding learned for each tokenizer head corresponding to an SSL layer model and a separate final linear layer for each transformer head. The baseline transformer model has 6 layers in the encoder, 12 layers in the decoder, 4 attention heads, a model dimension of 512 and a feed-forward layer in 2048. 

% We train all models for at least 200 epochs and then evaluate them for speech quality and fidelity to the text. To evaluate speech quality, we use a regression model trained on the SOMOS\cite{somos} dataset from Samsung to estimate human quality ratings. The model achieves a system-level correlation coefficient of 0.9 with human ratings. For fidelity to the text, we evaluate the generated samples using a pretrained ASR model and then compute the Word Error Rate and Character Error Rate.

\section{Results}
\label{sec:results}
\subsection{Scalable Vocoder}

% \begin{table*}[!h]
%  \centering\small
%  \caption{Scalable Vocoder.}  \label{tab:vocoder}
% \resizebox{\textwidth}{!}{
% \begin{tabular}{c|c|cc|cc|cc|cc|cc|cc}
%     \toprule
%       \multirow{3}{*}{SSL Model}& \multirow{3}{*}{Tokenizer} 
%       &\multicolumn{12}{c}{Layer (LJSpeech/LibriSpeech)} \\
%        \cmidrule(lr){3-14}
% & &\multicolumn{2}{c|}{3}&\multicolumn{2}{c|}{7}&\multicolumn{2}{c|}{12}&\multicolumn{2}{c|}{18}&\multicolumn{2}{c|}{23}&\multicolumn{2}{c}{All}\\
% \cmidrule(lr){3-4}\cmidrule(lr){5-6}\cmidrule(lr){7-8}\cmidrule(lr){9-10}\cmidrule(lr){11-12}\cmidrule(lr){13-14}
% & & MOS$\uparrow$& DWER$\downarrow$ & MOS$\uparrow$& DWER$\downarrow$& MOS$\uparrow$& DWER$\downarrow$& MOS$\uparrow$& DWER$\downarrow$& MOS$\uparrow$& DWER$\downarrow$& MOS$\uparrow$& DWER$\downarrow$\\
% \midrule
%     \multirow{2}{*}{Hubert Large\cite{hsu2021hubert}}&LJSpeech& 0 &0 &0 & 0& 0& 0& 0&0 & 0 & 0 &0 &0 \\
%       & LibriSpeech & 2.69-2.53 & 43.71-72.61 & 3.23-3.09  & 6.03-13.57 & 2.97-2.85 & 6.19-12.57& 2.44-2.35 & 25.88-37.64 & 2.83-2.58 & 5.69-14.22 & 3.48-3.41 & 2.92-7.15 \\
%       \midrule
%           \multirow{2}{*}{WavLM Large\cite{chen2022wavlm}}&LJSpeech& 0&0 &0 & 0& 0& 0& 0&0 & 0 & 0 &0 &0 \\
%       & LibriSpeech & 0&0 &0 & 0& 0& 0& 0&0 & 0 & 0 &0 &0 \\
%       \midrule
% \end{tabular}}
% \end{table*}
\begin{figure}[t]
  \centering
  \includegraphics[width=0.6\linewidth]{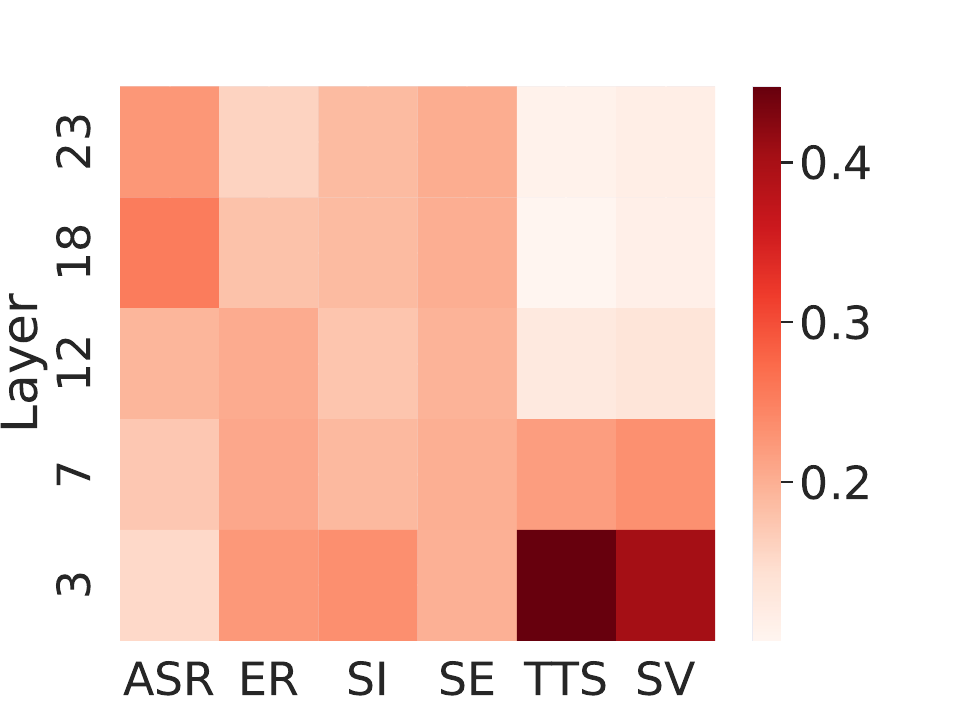}
  \caption{Attention analysis across various tasks and layers of the discrete WavLM model with in-domain tokenizers.}
\label{fig:att_analysis}
\end{figure}

\begin{table*}[t]
 \centering\tiny
\caption{Assessing the impact of the number of clusters and embedding initialization on discrete WavLM-Large across different tasks.}
 \label{tab:setting}
 \scalebox{0.60}{
\resizebox{\textwidth}{!}{
\begin{tabular}{c|c|c|c|c|cc|cc}
    \toprule
       \multirow{2}{*}{\textbf{Setting}}
      & \textbf{ASR (EN)} &  \textbf{ASR (FR)}  & \textbf{SID}& \textbf{ER} &\multicolumn{2}{c|}{\textbf{SE}}&  \multicolumn{2}{c}{\textbf{TTS}}\\
       \cmidrule(lr){2-2}
       \cmidrule(lr){3-3}\cmidrule(lr){4-4} \cmidrule(lr){5-5}\cmidrule(lr){6-7}\cmidrule(lr){8-9}
    &  \textit{WER} $\downarrow$ & \textit{WER} $\downarrow$ & \textit{ACC} $\uparrow$ & \textit{ACC} $\uparrow$&\textit{DNSMOS}$\uparrow$ & \textit{dWER}$\downarrow$\  & \textit{UTMOS}$\uparrow$\  & \textit{WER} $\downarrow$\ \\
     \midrule
       \multicolumn{9}{c}{Effect of Number Of Clusters}\\
         \midrule
1000 & 7.15 &34.61 &79.0 &61.8 & 3.93& 6.75& 3.65 &  5.76 \\
2000 & 6.96 &32.94 &79.5 &67.2 & 3.93 &6.58 &3.55 &5.62
 \\
     \midrule

       \multicolumn{9}{c}{Effect of Embedding Initialization}\\
            \midrule
Random & 6.96 &32.94 &81.0 &67.2 & 3.93& 6.75&3.65 &  5.76 \\
PreTrained \& finetune & 8.93 & 35.81 &77.5 &63.9  & 3.93 &6.82 & 3.64 &6.62\\
PreTrained \& freeze & 9.26 & 35.12 & 73.1 &67.0 & 3.93 &6.98 & 3.66 &6.42 \\
\bottomrule
\end{tabular}}}
\end{table*}

\begin{table*}[!h]
 \centering\tiny
 \caption{Out-of-domain and in-domain performance of discrete HuBERT and WavLM models across the downstream tasks.}
  \label{tab:result}
\scalebox{0.75}{
\resizebox{\textwidth}{!}{
\begin{tabular}{c|c|c|c|c|c|cc|cc|cc}
    \toprule
      \multirow{2}{*}{\textbf{SSL Model}}& \multirow{2}{*}{\textbf{Tokenizer}} 
      & \textbf{ASR (EN)} &  \textbf{ASR (FR)}  & \textbf{SID}& \textbf{ER} &\multicolumn{2}{c|}{\textbf{SE}}&  \multicolumn{2}{c|}{\textbf{TTS}} &\multicolumn{2}{c}{\textbf{Vocoder}}\\
       \cmidrule(lr){3-3}\cmidrule(lr){4-4}\cmidrule(lr){5-5} \cmidrule(lr){6-6}\cmidrule(lr){7-8}\cmidrule(lr){9-10}\cmidrule(lr){11-12}
    & &  \textit{WER} $\downarrow$ & \textit{WER} $\downarrow$ & \textit{ACC} $\uparrow$ & \textit{ACC} $\uparrow$ & \textit{DNSMOS}$\uparrow$ & \textit{dWER}$\downarrow$\   & \textit{UTMOS}$\uparrow$ & \textit{WER} $\downarrow$ & \textit{UTMOS}$\uparrow$ & \textit{dWER}$\downarrow$\ \\
     \midrule
      \multirow{2}{*}{HuBERT Large\cite{hsu2021hubert}}&In-Domain& 7.89&38.29
 &67.2 & 64.5& 3.98& 17.64&3.61 &  6.46 & 3.50
 & 4.49
\\
      & Out-Of-Domain & N/A&39.50 & 67.8 & 61.7& 3.95&15.92&3.54 & 5.45 & 3.48
 & 2.92 \\
    \midrule
    \multirow{3}{*}{WavLM Large\cite{chen2022wavlm}}&In-Domain& 6.96&32.94& 81.0 & 67.2&  3.93& 6.75&  3.65 &  5.76 & 3.49
 & 2.98
\\
      & Out-Of-Domain &N/A &36.25 &79.0 & 61.9&  3.96&6.49& 3.61 & 5.73& 3.68
 & 2.95
 \\

\bottomrule
\end{tabular}}}
\end{table*}
Our results cover findings from two distinct setups: 1) a scalable vocoder trained across five layers, and 2) a vocoder trained on a single layer. In both setups, the tokenizers and the vocoders are trained with LJSpeech (in-domain condition). In the first scenario, models are trained with HuBERT discrete tokens and WavLM discrete tokens, each with the number of clusters set to 1000. To further explore the influence of k-means cluster size on speech quality, we introduce an additional model with the number of clusters set to 2000. In the second setup, we focus on models trained specifically on a single layer (3 or 23) using HuBERT discrete tokens and in-domain tokenizer. This experiment aims to compare the performance of the scalable vocoder against the vocoder trained on a single layer.

The results are summarized in Fig.~\ref{fig:imbalance}.  WavLM combined with an in-domain tokenizer achieves higher UTMOS and lower dWER scores across all setups. About the impact of the number of clusters, our experiment shows that setting $k$ to 2000 degrades the quality of synthesized speech. Finally, both models trained on a single layer are outperformed on both evaluation metrics by the one trained on five layers, confirming the benefits of the scalable approach.
Lastly, we explore an out-of-domain scenario where the tokenizers are trained on LibriSpeech and the vocoders are on LJSpeech. As shown in Table~\ref{tab:result} (last column), we do not observe any significant performance degradation when using the scalable vocoder in an out-of-domain condition.

\subsection{Layer Analysis}
Fig. \ref{fig:att_analysis} depicts the average weights assigned to different layers in the WavLM model across various downstream tasks on the test dataset.  In both TTS and the scalable vocoder, lower levels get greater importance as they prioritize effective reconstruction. Conversely, for ASR, the upper layers become more crucial in capturing the semantic aspects of spoken utterances. In the case of ER and SID, the third layer receives the highest weight. Our findings align with the observed pattern in continuous representations \cite{Zaiem2023speech2}. For SE, all layers are equally weighted, indicating the necessity of all hierarchical levels to achieve optimal audio quality while preserving the semantic content of the input. 
% \begin{figure}[t]
%   \centering
%   \includegraphics[width=\linewidth]{figures/examples/example1.png}
%   \caption{Attention analysis on different layers of Hubert and WavLM models across different tasks. We adopt the similar configuration used in the In-Domain experiments in Table \ref{tab:result}.}
% \label{fig:att_analysis}
% \end{figure}

% \begin{figure}[!t]
%     \centering
%     \begin{subfloat}{}
%     \includegraphics[width=1.5in]{figures/wavlm_att.pdf}
%     \end{subfloat}
%     \begin{subfloat}{}
%     \includegraphics[width=1.5in]{figures/hubert_att.pdf}
%     \end{subfloat}
% \caption{Attention analysis on different layers of WavLM model across different tasks. We adopt the similar configuration used in the in-domain experiments in  Table \ref{tab:result}.}
% \label{fig:att_analysis}
% \end{figure}

% \begin{table}[!h]
%  \centering
%  \caption{ Comparison of leveraging many vs one  SSL layer on re-synthesized speech in preserving audio information. }  \label{tab:single_vs_multi}
% \begin{tabular}{cccc}
% \toprule
% Model & WER  & UTMOS  \\
% \midrule
% Vocoder with Bitrate Scalability & 0  & 0 \\
% \midrule
% Vocoder on Single Layer & 0 & 0 \\
% \end{tabular}
% \end{table}

\subsection{Effect of Number of Clusters}
\label{subsec:cluster}
 We train k-means models with both 1000 and 2000 centroids and examine the impact of the number of clusters across different tasks, as illustrated in Table \ref{tab:setting}.  In Generative tasks, TTS and SE, no significant differences are observed between models trained with 1000 and 2000 clusters. However, for ASR in both English and French, as well as ER, models with a higher number of clusters outperform those with fewer clusters. In the case of SID, the model trained with 1000 clusters exhibits comparable accuracy to the model with 2000 centroids. As expected, the ideal number of clusters is task-dependent. For multi-modal LLMs where a single set of tokens is desired to solve multiple tasks,  we recommend a cluster count between 1000 and 2000.

\subsection{Effect of Embedding Initialization}
\label{subsec:init}
We study various configurations for initializing the embedding layers of audio tokens (Table \ref{tab:setting}). Three options are considered: \begin{enumerate*}[label={\arabic*)}]
  \item Random initialization of the embedding layers,
  \item  Initialization of the embedding layer with the corresponding centroid's embedding, while freezing the layer, and 
  \item Initialization of the embedding layer with the corresponding centroid's embedding, without freezing the layer.
\end{enumerate*} Across all tasks, there is no advantage observed in initializing the embedding with pretrained centroid embeddings, and random initialization consistently outperforms it in all scenarios. However, discriminative tasks show greater benefits from random initialization, while generative tasks exhibit comparable performance across all three settings. This observation eliminates the need for having the same embedding size as the SSL models, allowing the choice of a smaller and more efficient embeddings. %In all tasks, except for ER, achieving better or comparable performance is possible when fine-tuning the pretrained embeddings rather than freezing them. 
% \begin{table*}[!h]
%  \centering\small
%  \caption{Evaluating effect of embedding initialization on various downstream tasks.}  \label{tab:init}
% \resizebox{\textwidth}{!}{
% \begin{tabular}{c|c|c|c|c|ccc|cc}
%     \toprule
%        \multirow{2}{*}{Initialization Method}
%       & ASR &\multicolumn{1}{p{2cm}|}{\centering Multilingual \\ ASR}   & SID& ER &\multicolumn{3}{c|}{SE}&  \multicolumn{2}{c}{TTS}\\
%        \cmidrule(lr){2-2}
%        \cmidrule(lr){3-3}\cmidrule(lr){4-4} \cmidrule(lr){5-5}\cmidrule(lr){6-8}\cmidrule(lr){9-10}
%     &  WER $\downarrow$ & WER $\downarrow$ & EER $\downarrow$ & ACC $\uparrow$& SI-SNR$\uparrow$ & DNSMOS$\uparrow$ & DWER$\downarrow$\  & WER $\downarrow$ & MOS$\uparrow$\\
%      \midrule
% random & 0 &0 &0 &0 &0 & 0 &0 &0 &0 \\
% random \& finetune & 0 &0 &0 &0 &0 & 0 &0 &0 &0 \\
% random \& freeze & 0 &0 &0 &0 &0 & 0 &0 &0 &0 \\
% \end{tabular}}
% \end{table*}

\subsection{Out-of-Distribution Generalization}
To evaluate the robustness of discrete representations under distribution shifts, we train tokenizers on both in-domain and out-of-domain datasets (Table \ref{tab:result}). In discriminative tasks, k-means models are trained using the same dataset employed for training acoustic models. In the out-of-domain scenarios, k-means models are trained on \textit{train-clean-100}, \textit{train-clean-360}, and \textit{train-other-500}. For generative tasks, k-means models are trained on LJSpeech for in-domain evaluation and \textit{LibriSpeech-960h} for OOD evaluation, while both the acoustic model and vocoder are trained on LJSpeech. For all discriminative tasks, the in-domain tokenizer outperforms its OOD counterpart. Interestingly, in all generative tasks, training the model using the OOD tokenizer does not adversely affect performance and, in some instances, even improves the results. We speculate that this trend may arise because generative tasks primarily depend on tokens capturing low-level information, which tends to be more ``universal" and transferable across different domains.

\section{Conclusions}
Discrete semantic tokens, derived from the quantization of SSL models, play an important role, providing  \textit{``pseudo-text"} valuable for training text-free speech language models and multimodal LLMs. We explore the optimal configuration of semantic tokens across discriminative and generative tasks. We introduce a novel technique involving an informed layer selection mechanism, utilizing learnable attention weights to integrate information from different SSL layers. This approach significantly enhances the performance and interpretability of the model. Furthermore, we propose a scalable solution for training a universal vocoder across multiple SSL layers, demonstrating its superiority over vocoders trained on specific layers. As future work, we plan to explore more diverse tasks and quantization methods, and the development of a multi-speaker vocoder.
% By releasing code, we aim to encourage further research in the field. As future work, we plan to explore more diverse tasks and quantization methods, and the development of a multi-speaker vocoder.
% \section{Acknowledgements}
% We acknowledge the support of the Natural Sciences and Engineering Research Council of Canada (NSERC) and the Digital Research Alliance of Canada (alliancecan.ca).

% \ifinterspeechfinal
%      The Interspeech 2024 organisers
% \else
%      The authors
% \fi

\bibliographystyle{IEEEtran}

\bibliography{bibliography}

\end{document}